\documentclass[aps,
prd,
reprint,
amsmath,
amssymb,
amsthm, 
bm,
]{revtex4-2}

\usepackage{graphicx}
\usepackage{dcolumn}
\usepackage{multirow} 
\usepackage{empheq}
\usepackage{xcolor}
\usepackage{soul}
\usepackage{hyperref}
\usepackage{placeins} 
\usepackage{bm} 
\usepackage{amsmath}

\begin{document}


\title{Model-independent time-delay interferometry based on principal component analysis}



\author{Quentin Baghi}
\email[]{quentin.baghi@cea.fr}

\affiliation{IRFU, CEA, Universit\'{e} Paris-Saclay, F-91191 Gif-sur-Yvette, France}

\author{John Baker}
\author{Jacob Slutsky}
\author{James Ira Thorpe}

\affiliation{Goddard Space Flight Center \\
	Mail Code 663\\
	8800 Greenbelt Rd, Greenbelt, 
	Maryland 20771, USA}

\renewcommand\hl[1]{#1}


\date{\today}

\begin{abstract}
With a laser interferometric gravitational-wave detector in separate free flying spacecraft, the only way to achieve detection is to mitigate the dominant noise arising from the frequency fluctuations of the lasers via postprocessing. The noise can be effectively filtered out on the ground through a specific technique called time-delay interferometry (TDI), which relies on the measurements of time-delays between spacecraft and careful modeling of how laser noise enters the interferometric data. Recently, this technique has been recast into a matrix-based formalism by several authors, offering a different perspective on TDI, particularly by relating it to principal component analysis (PCA). In this work, we demonstrate that we can cancel laser frequency noise by directly applying PCA to a set of shifted data samples, without any prior knowledge of the relationship between single-link measurements and noise, nor time-delays. We show that this fully data-driven algorithm achieves a gravitational-wave sensitivity similar to classic TDI.
\end{abstract}


\maketitle


\section{\label{sec:intro}Introduction}

Gravitational waves can be detected through a constellation of satellites forming a network of laser interferometers that probe phase differences in beams exchanged between spacecraft. This concept is the core of future space-based detectors like the Laser Interferometer Space Antenna (LISA)~\cite{danzmann_lisa_2017} and TianQuin~\cite{milyukov_tianqin_2020}. It was soon realized~\cite{tinto_cancellation_1999} that the noise caused by the stochastic fluctuations of laser frequencies cannot be canceled at the photometric detectors onboard the satellites. The reason lies in the difference between the delays they experience when being transmitted along two interferometric arms. Rather, interferometry must be performed once the telemetry is retrieved on the ground. By taking advantage of the fact that several delayed versions of the same noise component appear in different measurements, it is possible to cancel laser noise by combining and interpolating the received data streams at the right timestamps. We call this post-processing technique time-delay interferometry (TDI)~\cite{tinto_time-delay_2014}. TDI has been undergoing a constant development to account for increasing degrees of complexity, including optical bench motion noise~\cite{estabrook_time-delay_2000}, varying armlengths~\cite{cornish_effects_2003, shaddock_data_nodate}, noise correlations~\cite{prince_lisa_2002}, on-board filtering~\cite{bayle_effect_2019}, clock jitters~\cite{hartwig_clock-jitter_2020} and the preferred representative units of the processed data (phase or fractional-frequency)~\cite{bayle_adapting_2021}. Other efforts focused on finding new noise-canceling combinations~\cite{Muratore2020}.

Romano and Woan~\cite{romano_principal_2006} showed that we could think of TDI combinations as low-variance principal components of the laser noise covariance matrix. This idea relied on a matrix representation of the interferometric measurements and was also explored in Ref.~\cite{leighton_principal_2016}. Computing TDI variables is equivalent to projecting the data onto the null space of the covariance. Recently, this concept regained interest in the community. Vallisneri \textit{et al.}~\cite{vallisneri_tdi-infinity_2020} presented a similar algebraic computation (called TDI-$\infty$) where they perform the singular value decomposition (SVD) of the design matrix $\bm{M}$, i.e., the matrix translating individual laser noise sources $\bm{c}$ into measured noises $\bm{y}$ so that $\bm{y} = \bm{M} \bm{c}$. They approximate the model likelihood by restricting it to the dominating terms, using the singular vectors that correspond to the null space of $\bm{M}$. Concomitantly, we derived a frequency-domain version of the covariance eigenvector decomposition~\cite{baghi_statistical_2021} called PCI (for principal component interferometry), where we focused on the orthogonalization of the data with respect to secondary (i.e., non-laser) noises. We showed that this method generalizes the A, E, and T TDI variables~\cite{prince_lisa_2002} and can bring an improvement in signal-to-noise ratio (SNR) when the levels of the noises affecting the single-link measurements are unequal. Although the implementation details are still under debate~\cite{tinto_matrix_2021, bayle_matrix_2021, tinto_reply_2021}, the matrix representation of TDI is a powerful tool for data analysis, as it allows one to express the model likelihood directly in terms of single-link measurements and readily account for data gaps.

TDI and the extensions cited above rely on accurate modeling of the dependence of the measurements on the various laser noise sources. TDI stems from a set of equations modeling every interferometer, from which one derives the sequence of combination and delaying operations needed to precisely cancel laser noise terms. Equivalently, TDI-$\infty$ assumes perfect knowledge of the design matrix, while PCI assumes a model of the noise covariance. In addition, these methods need precise estimates of the time-delays, which can be obtained independently from the telemetry~\cite{wang_first_2014} or directly estimated from the data using noise power minimization~\cite{tinto_time-delay_2005} or, as it was recently shown, by Bayesian inference~\cite{page_bayesian_2021-1, baghi_statistical_2021}.
However, LISA is a complex apparatus involving many control laws, couplings, and frequency planning inducing correlations among lasers. Inaccurate modeling can lead to imperfect  cancellation or even no laser noise suppression at all~\cite{bayle_effect_2019, hartwig_clock-jitter_2020}. In this work, we present a fully data-driven approach to cancel laser noise in space interferometry, with no prior assumption on the noise structure. The method (that we name aPCI for automatic PCI) relies on the principal component analysis (PCA) of a matrix directly constructed from the measurements.

PCA is usually used to reduce the dimensionality of a problem by decomposing the data into an orthogonal basis whose vectors are called principal components. Then, one keeps only the vectors that contain the most information about the data by restricting the analysis to the components having the largest variance. However, when the noise overwhelmingly dominates the signal of interest, as this is the case for gravitational waves (GW) by about a factor $10^8$, we can use PCA with the opposite objective: to retain the components having the lowest variance. 

In most problems, the decomposition is performed on a collection of $n$ different realizations of the same $p$-dimensional measurement, forming a matrix of size $n \times p$. However, in a gravitational-wave observatory like LISA, the data is instead made of different channels affected by different noises and measuring different projections of the same signal. To build a data matrix that we can process with PCA, we can draw inspiration from the technique used to form the TDI combinations~\cite{shaddock_postprocessed_2004}, which relies on interpolators like Lagrange filters to produce delayed versions of the measured data. The measurement estimate at a given point in time is interpolated by linearly combining nearby data samples. Thus, we build a $n \times p$ data matrix whose rows are versions of the same data shifted by an integer number of samples. Then, we compute the PCA of this matrix to obtain components organized in decreasing order of variance. The lowest variance components correspond to combinations where laser noise is significantly reduced. Projecting the data onto these components yields data streams that are readily usable for gravitational-wave searches.

We first detail the principle of the aPCI method in Section~\ref{sec:pca}, and present the theoretical framework. In Section~\ref{sec:evaluation}, we provide a means to evaluate the method performance by deriving the sensitivity of aPCI combinations to monochromatic GWs in the case of LISA. In Section~\ref{sec:simulations}, we use a numerical simulation of LISA data to demonstrate laser noise suppression, and also compare aPCI with classic TDI by computing their sensitivities averaged over sky location and orientation.  
Finally, in Section~\ref{sec:discussion} we discuss important aspects of this approach and pave the way for future studies.

\section{\label{sec:pca}Principal component analysis theoretical framework}

\subsection{\label{sec:data_matrix}Construction of the data matrix}

Space-based gravitational-wave detector data include the outputs of several interferometers, which can be expressed in relative frequency deviations. We denote the corresponding measurements as $s_{ij}(t)$ to conform to the official convention (see, for example, Ref.~\cite{bayle_adapting_2021}), where $i$ is the index of the satellite hosting the optical bench, and $j$ is the index of the sending spacecraft. 
In practice, we measure a discrete version of $s_{ij}(t)$ with $N$ samples spaced every $\tau_s$ seconds. Therefore, we can form a vector $\bm{s}_{ij}$ with $N$ entries given by $s_{ij}(t_n)$, where the time stamps are $t_{n} = n \tau_s, \, n \in \left[ 0, \, N-1\right]$. Hence, we have $\bm{s}_{ij} \equiv \left(s_{ij}(t_0),\, \hdots,\,  s_{ij}(t_{N-1})\right)^{\dagger}$. We will stick to the LISA example, where the detector is made of a constellation of 3 satellites, each of them carrying 2 optical benches on board. If we consider the science measurements only, then we have 6 measurements that we can stack into a $N \times 6$ matrix $\bm{Y}$ as
\begin{align}
	\label{eq:y_matrix}
	\bm{Y} \equiv \left(\bm{s}_{12}, \, \bm{s}_{23}, \, \bm{s}_{31}, \, \bm{s}_{13}, \, \bm{s}_{21}, \, \bm{s}_{32} \right).
\end{align}

Classical TDI algorithms usually operate in two steps. First, they compute delayed versions of discrete signals, interpolated at specific times depending on the light travel time delays along the constellation arms. In a second step, they combine these interpolated time series in such a way that laser noise terms vanish.

We would like to form laser-noise canceling combinations of the data without knowing the delays nor the way they enter the measurements. We know that the first step of TDI is usually performed through fractional-delay filter techniques which use linear combinations of samples at times before and after the target date~\cite{shaddock_postprocessed_2004}. In this process, one must choose the number of samples $n_h$ to use before and after the desired interpolation time. In principle, the larger $n_h$, the better the interpolation accuracy, but the longer the computation time. To provide a fully data-driven algorithm with the ingredients necessary to automatically find the correlations among the data, we can therefore start from shifted versions of the measurement vector $\bm{s}$. Let us assume that we have measured data over a time span from $t=0$ to $t = T_{\mathrm{obs}}$, sampled every $\tau_s$. We can build a matrix $\bm{X}$ containing $\bm{Y}$ itself, along with $n_h$ backward-shifted versions of it and $n_h$ forward-shifted versions, resulting in a matrix of size $N \times 6 (2n_h + 1) $ such that:
\begin{align}
	\label{eq:data_matrix}
	\bm{X} \equiv \left( \bm{D}_{-n_h}\bm{Y},\, \bm{D}_{-n_h+1}\bm{Y}, \, \hdots,  \, \bm{D}_{n_h}\bm{Y} \right),
\end{align}
where we defined the integer delay matrix $\bm{D}_{m}$ of size $N \times N$ which acts on any vector $\bm{s}$ as
\[{\bm{D}_{m} \bm{s}}_{n}  = \begin{cases} 
	\bm{s}_{n-m} &  \text{if } n \in \left[n_{\mathrm{min}}, \, n_{\mathrm{max}}\right]\\
	0 & \text{ otherwise, }
\end{cases}
\]
with $n_{\mathrm{min}} \equiv \mathrm{max}(0, m)$ and $n_{\mathrm{max}} \equiv \mathrm{min}(N-1, N-1+m)$. 
This way, the matrix $\bm{X}$ is formed of sub-vectors including 6 interferometric measurements, where a sub-vector is shifted by one time sample with respect to the previous one.
In this form, the data is ready to be processed through PCA. In the following, we will set $p \equiv 2 n_h + 1$ as the number of subchannels. 

\subsection{\label{sec:pca_decomposition}PCA decomposition of the data matrix}

PCA aims at finding a transformation of the data matrix into a new set of variables which are as independent as possible with respect of the information they carry, measured in terms of variance. This is done by computing a set of vectors called principal components (PCs) representing orthogonal directions that are ordered from the one that best fits the data, to the one that is less representative. In the problem we study, the data matrix variance overwhelmingly represents laser noise. The first PCs will therefore better fit the laser noise. On the contrary, the last PCs should be more independent of laser noise, representing the remaining information present in the data.

We can compute the PCs through singular value decomposition (SVD):
\begin{equation}
\label{eq:svd}
	\bm{X} = \bm{U} \bm{S} \bm{V}^{\intercal},
\end{equation}
where $\bm{U}$ is a $N \times N$ unitary matrix of left-singular vectors, $\bm{S}$ is a $N \times 6p$ rectangular diagonal matrix whose non-zero elements are the singular values, $\bm{V}$ is a $6p \times 6p$ unitary matrix of right-singular vectors, and $\intercal$ is the matrix transpose operator. The PCs are given by the vectors forming the columns of $\bm{V}$.

Note that, in regular PCA, the columns of the matrix we decompose correspond to different features (like different sensor outputs) and the rows correspond to different realizations of the measurement. In the way we apply it to laser noise processing, the matrix columns include not only different interferometer outputs, but also different shifts, while the rows correspond to particular times at which the data is measured. Therefore, to draw a comparison between TDI and what we do here, the elements of particular PC vector could be interpreted as the kernels of several fractional delay filters.

The principal component decomposition of $\bm{X}$ is therefore the matrix $\bm{T}$ resulting in the transformation
\begin{equation}
	\label{eq:pca_transformation}
	\bm{T} \equiv \bm{X} \bm{V}.
\end{equation}
The rows of $\bm{T}$ are usually called the scores, but we will refer to them as the PCA (or aPCI) combinations in the following.

In this problem, we are interested in the $q$ lowest-variance components of the decomposition, below some threshold where the laser noise is significantly suppressed. Let us rewrite $\bm{V}$ in two parts: a part $\bm{V}_q=\bm V\bm{\Pi}_q$, where $\bm{\Pi}_q \equiv \left(\bm{0}, \, \bm{I}_{q} \right)^{\intercal}$ is the $6p \times q$ matrix that projects the last $q$ low-variance PCs, and the complementary part $\bm{V}_{\setminus{q}}$ containing the first $2p-q$ PCs related to large variance: 
\begin{equation}
\label{eq:pca_transformation_truncated}
	\begin{pmatrix} \bm{T}_{\setminus{q}} \\ \bm{T}_q \end{pmatrix} \equiv \bm{X} \begin{pmatrix}\bm{V}_{\setminus{q}}& \bm{V}_q \end{pmatrix},
\end{equation}
where we split $\bm{V} = \left(\bm{V}_{\setminus{q}}, \, \bm{V}_q \right)$. To a good approximation, the low-variance transformation $\bm{T}_{q}$ should be all we need to search for the signals of interest, i.e. gravitational waves. 

A key effect of the PCA is that the transformation from $\bm{X}$ to $\bm{T}$ diagonalizes the data matrix sample covariance $\bm{\hat{\Sigma}}_{\bm{X}}=\frac1N\bm{X}^\intercal\bm{X}$ and  transforms it to the diagonal matrix  $\bm{\hat{\Sigma}}_{\bm{T}}=\frac1N\bm{T}^\intercal\bm{T}=\frac1N\bm{S}^\intercal\bm{S}$. Here we are primarily interested in the \emph{inverse} covariance, which the low-variance decomposition allows us to write as:
\begin{eqnarray}
\label{qe:sigma-t-inv-approx}
	\bm{\hat\Sigma}^{-1}_{\bm{T}} &=& \begin{pmatrix} \bm{V}_{\setminus{q}}^{\intercal}\bm{\hat\Sigma}^{-1}_{\bm{X}}\bm{V}_{\setminus{q}}
	& \bm{V}_{\setminus{q}}^{\intercal}\bm{\hat\Sigma}^{-1}_{\bm{X}}\bm{V}_{q}\\
      \bm{V}_{q}^{\intercal}\bm{\hat\Sigma}^{-1}_{\bm{X}}\bm{V}_{\setminus{q}} & \bm{V}_{q}^{\intercal}\bm{\hat\Sigma}^{-1}_{\bm{X}}\bm{V}_{q} \end{pmatrix} \nonumber \\
      & \approx & \begin{pmatrix} \bm{0} & \bm{0} \\
      \bm{0} & \bm{V}_{q}^{\intercal}\bm{\hat\Sigma}^{-1}_{\bm{X}}\bm{V}_{q} \end{pmatrix} \nonumber  \\
      & = & \bm{\Pi}^{\intercal}_q \bm{\Pi}_q \bm{\hat\Sigma}^{-1}_{\bm{T}} \bm{\Pi}^{\intercal}_q \bm{\Pi}_q
\end{eqnarray}
where in the last line we used the diagonality of $\bm{\hat{\Sigma}}_{\bm{T}}$ and neglected the first $6p-q$ inverse variances compared to the $q$ last ones. Here our reverse-ordered preference for the nominally least-significant principal components has the effect of selecting the most significant parts of the inverse covariance matrix.

\subsection{\label{sec:covariance}Relation of the PCA process with the model covariance}

As is common in gravitational-wave data analysis, we assume our noise is Gaussian and stationary (over the relatively short timescales relevant for laser noise.) Stationarity allows us to assume that frequency bins are approximately uncorrelated in the Fourier domain. For any vector $\bm{z}$, we will label as $\bm{\tilde{z}}(f)$ its discrete Fourier transform (DFT) element at frequency $f$. Similarly, we will denote by $\bm{\tilde{\Sigma}}_{z}(f)$ its Fourier-domain covariance, which fully characterizes the (zero-mean) noise.

The construction of the data matrix in Eq.~\eqref{eq:data_matrix} can be rewritten in the frequency domain as 
\begin{equation}
    \label{eq:data_matrix_fourier}
    \bm{\tilde{X}}(f) = \bm{\tilde{y}}^{\intercal}(f) \begin{pmatrix} \bm{\tilde{D}}_{-n_h}(f), \, \hdots , \, \bm{\tilde{D}}_{+n_h}(f)  \end{pmatrix} ,
\end{equation}
where $\bm{\tilde{y}}(f)$ is now a $6 \times 1$ column vector, and $\bm{\tilde{X}}(f)$ is a $1 \times 6p$ row vector. Here we approximate the $m^{\mathrm{th}}$ integer delay operator involved in Eq.~(\ref{eq:data_matrix}) by a multiplicative complex exponential phase in the Fourier domain, encoded by a $6 \times 6$ diagonal matrix $\bm{\tilde{D}}_{m}(f)$ with entries
\begin{equation}
	\tilde{D}_{m, pq}(f) = \delta_{pq} e^{-2 i \pi f m \tau_s},
\end{equation}
where $i = \sqrt{-1}$ is the complex number and $\delta_{pq}$ is the Kronecker symbol ($\delta_{pq} = 1 \text{ if } p = q$ and $\delta_{pq} = 0$ otherwise). 

Similarly, we can write a frequency-domain version of the projection in  Eq.~\eqref{eq:pca_transformation} by defining $\bm{\tilde{t}}(f)$ as the $6p \times 1$ vector formed by stacking the $6p$ PCA combinations at frequency $f$. Then, we have
\begin{align}
    \label{eq:pca_transformation_fourier}
	\bm{\tilde{t}}(f) &= \bm{V}^{\intercal} \bm{\tilde{X}}^{\intercal}(f) = \bm{W}(f) \bm{\tilde{y}}(f), 
\end{align}
where we set $\bm{\tilde{W}}(f) \equiv \bm{V}^{\intercal} \big( \bm{\tilde{D}}_{-n_h}(f), \, \hdots , \, \bm{\tilde{D}}_{+n_h}(f) \big)^{\intercal}$ the $6p \times 6$ matrix transformation which encodes, in the Fourier domain, the construction of the data matrix and its PCA transformation. 

The transformation matrix $\bm{\tilde{W}}$ has useful properties:
\begin{eqnarray}
\label{eq:w_unitary}
    \bm{\tilde{W}}^{\dag} \bm{\tilde{W}} &=& \begin{pmatrix} \bm{\tilde{D}}_{-n_h}^{\dag} & \hdots &  \bm{\tilde{D}}_{+n_h}^{\dag} \end{pmatrix} \bm{V} \bm{V}^{\intercal} \begin{pmatrix}\bm{\tilde{D}}_{-n_h} \\ \vdots \\ \bm{\tilde{D}}_{+n_h} \end{pmatrix} \nonumber \\
    & = & \sum_{m=-n_h}^{+n_h} \bm{\tilde{D}}_{m}^{\dagger}\bm{\tilde{D}}_{m} = p \bm{I}_{6},
\end{eqnarray}
where we used the fact that $\bm{V}$ is unitary. Note that here we took the Hermitian transpose $\dag$, because $\bm{\tilde{W}}$ is a complex-valued matrix. 

We denote by $\bm{\tilde{\Sigma}}_{\bm{x}} \equiv \operatorname{E}\left[\bm{\tilde{x}} {\bm{\tilde{x}}}^{\dagger} \right]$ the covariance of any zero-mean complex random vector $\bm{\tilde{x}}$. Then from Eq.~\eqref{eq:pca_transformation_fourier} we can write the $6p \times 6p$ covariance of $\bm{\tilde{t}}$ as 
\begin{equation}
	\label{eq:pca_channels_covariance_fourier}
	\bm{\tilde{\Sigma}}_{\bm{t}} = \bm{\tilde{W}} \bm{\tilde{\Sigma}}_{\bm{y}} \bm{\tilde{W}}^{\dagger},
\end{equation}
where we dropped the dependence on $f$ to lighten the expressions. 

We seek a similar expression relating the inverse covariances. Because the transformation $\bm{\tilde{t}}=(1/p)\bm{\tilde{W}}^\dag \bm{\tilde{y}}$ involves a projection from $6p$ dimensions down to $6$, we do not have a useful inverse identity involving $\bm{\tilde{W}}\bm{\tilde{W}}^\dag$.  Consequently it is not straightforward to find an expression for the inverse covariance by solving $\bm{A} \bm{\tilde{\Sigma}}_{\bm t} = \bm{I}_{6p}$.  However, if we set $\bm{A}=(1/p^2)\bm{\tilde{W}}\bm{\Sigma}_{\bm{y}}^{-1}\bm{\tilde{W}}^\dag$, we can verify that $\bm{\tilde\Sigma}_{\bm t}\bm A\bm{\tilde\Sigma}_{\bm t}=\bm{\tilde\Sigma}_{\bm t}$ and  $\bm A \bm{\tilde\Sigma}_{\bm t}\bm A=\bm A$ and that both $\bm A \bm{\tilde\Sigma}_{\bm t}$ and $\bm{\tilde\Sigma}_{\bm t}\bm A $ are Hermitian, thus establishing that A is the pseudoinverse of $\bm{\tilde\Sigma}_{\bm t}$, which we simply denote as
\begin{equation}
    \bm{\Sigma}_{\bm t}^{+}\bm =(1/p^2)\bm{\tilde{W}}\bm{\Sigma}_{\bm y}^{-1}\bm{\tilde{W}}^\dag.
\end{equation}
Using the explicit properties of $\bm{\tilde{W}}$, we can then invert this expression to find
\begin{equation}
    \bm{\Sigma}_{\bm y}^{-1}=\bm{\tilde{W}}^\dag\bm{\Sigma}_{\bm t}^{+}\bm{\tilde{W}}.
\end{equation}

If the sample variance from the PCA data matrix provides a sufficient estimate of the (generally distinct) data variance as regards the laser noise, then as in Section~\ref{sec:pca_decomposition} we can estimate $\bm{\tilde{\Sigma}}_{\bm{t}}^{+} \approx \bm{\Pi}^{\intercal}_q \bm{\Pi}_q\bm{\tilde{\Sigma}}_{\bm{t}}^{+}\bm{\Pi}^{\intercal}_q \bm{\Pi}_q.$  Thus
 \begin{eqnarray}
 	\label{eq:inverse_covariance_approx}
 	\bm{\tilde{\Sigma}}_{\bm{y}}^{-1} &\approx& \bm{\tilde{W}}^{\dag}_{q} \bm{\Pi}_q \bm{\tilde{\Sigma}}_{\bm{t}}^{+}\bm{\Pi}^{\intercal}_q \bm{\tilde{W}}_{q}. 
 \end{eqnarray}   
where we set $\bm{\tilde{W}}_{q} \equiv \bm{\Pi}_q \bm{\tilde{W}}$.

This approximation amounts to ignoring the negligible terms depending on laser noise when computing the inverse covariance, through a suitable transformation that allows us to restrict the analysis to the terms that are relevant for the GW parameter inference.

\subsection{\label{sec:bayesian}Bayesian inference}

Gravitational-wave signals are usually analyzed using Bayesian inference.  
We consider the frequency-domain data $\boldsymbol{\tilde{y}}$ comprising six channels evaluated at frequency bin $f$. We drop the explicit dependence on $f$ for more clarity.  
For a model parameterized by $\boldsymbol{\theta}$, the posterior distribution given $\boldsymbol{\tilde{y}}$ is
\begin{eqnarray}
	\label{eq:posterior}
	p\left( \boldsymbol{\theta} | \boldsymbol{\tilde{y}} \right) = \frac{ p\left(\boldsymbol{\tilde{y}}  |  \boldsymbol{\theta} \right)  p\left(\boldsymbol{\theta} \right) }{ p\left(\boldsymbol{\tilde{y}}\right) },
\end{eqnarray}
where $p\left(\boldsymbol{\tilde{y}}  |  \boldsymbol{\theta} \right)$ is the likelihood, $p\left(\boldsymbol{\theta} \right)$ is the prior distribution of the parameters, and the evidence  $p\left(\boldsymbol{\tilde{y}}\right)$ provides normalization.

For zero-mean Gaussian noise, the likelihood has the form: 
\begin{eqnarray}
	\label{eq:likelihood}
	p\left( \boldsymbol{\tilde{y}}  |  \boldsymbol{\theta} \right) = \frac{\exp\left\{- \frac{1}{2} \left(\boldsymbol{\tilde{y}} -  \boldsymbol{\tilde{h}} \right)^{\dagger}\boldsymbol{\tilde{\Sigma}}_{\bm{y}}^{-1} \left(\boldsymbol{\tilde{y}} -  \boldsymbol{\tilde{h}} \right)\right\}}{\sqrt{(2 \pi)^{6}  \left| \boldsymbol{\tilde{\Sigma}}_{\bm{y}} \right| }},
\end{eqnarray}
where $\dagger$ denotes the Hermitian conjugate, $\boldsymbol{\tilde{h}}$ is the contribution to the data $\boldsymbol{\tilde{y}}$ expected from the gravitational wave signal implied by parameters $\boldsymbol{\theta}$, and $\boldsymbol{\tilde{\Sigma}}_{\bm{y}}$ is the $6N \times 6N$ covariance.
At this level the covariance includes all the noise processes including dominant contributions from the laser noise.

Bayesian analysis is clearest, when the data represented in the analysis comes directly from the basic measurements.  In practice though some pre-treatment or approximating assumptions can be important to make the analysis practical. For LISA, the challenge is that the basic problem represented in Eq.~\eqref{eq:likelihood} is dominated by the uninteresting laser noise.  TDI processing is one approach to recast the problem without this dominant part.  

Here we propose to apply our PCA treatment instead. To see how this works, we first plug the approximation for the inverse covariance derived in Eq.~\eqref{eq:inverse_covariance_approx} into Eq.~\eqref{eq:likelihood} and take the logarithm to obtain:
\begin{eqnarray}
	\label{eq:loglikelihood_approx}
	\ln p\left( \bm{\tilde{y}}  |  \boldsymbol{\theta} \right) &\approx& - \frac{1}{2} \left(\bm{\tilde{y}}-\bm{\tilde{h}} \right)^{\dag} \bm{\tilde{W}}^{\dagger}_{q}
	\bm{\Pi}_{q}\bm{\tilde{\Sigma}}_{\bm{t}}^{+}\bm{\Pi}^{\dag}_{q} 
	\bm{\tilde{W}}_{q} \left(\bm{\tilde{y}}-\bm{\tilde{h}}  \right) \nonumber \\ 
	&& - \frac{1}{2}\ln{\left((2 \pi)^{6} \big| \bm{\Pi}_q^{\dag} \bm{\tilde{W}} \bm{\tilde{\Sigma}}_{\bm{y}} \bm{\tilde{W}}^{\dagger} \bm{\Pi}_q \big|\right) }.
\end{eqnarray}

Then, we use the transformation in Eq.~\eqref{eq:pca_transformation_fourier} to write the log-likelihood as a function of $\bm{\tilde{t}}$:
\begin{eqnarray}
	\label{eq:loglikelihood_approx_t}
	\ln p\left( \bm{\tilde{y}}  |  \boldsymbol{\theta} \right) &=& - \frac{1}{2} \left(\bm{\tilde{t}}_q-\bm{\tilde{t}}_{hq} \right)^{\dag} \bm{\tilde{\Sigma}}_{\bm{t}_q\bm{t}_q}^{-1} \left(\bm{\tilde{t}}_{q}-\bm{\tilde{t}}_{hq}  \right) \nonumber\\ && - \frac{1}{2}\ln{\left((2 \pi)^{6} \big| \bm{\tilde{\Sigma}}_{\bm{t}_q\bm{t}_q} \big|\right) },
\end{eqnarray}
where we have also defined 
$\bm{\tilde{\Sigma}}_{\bm{t}_q\bm{t}_q}^{-1} \equiv \bm{\Pi}_{p} \bm{\tilde{\Sigma}}_{\bm{t}}^{+}\bm{\Pi}_{p}$, $\bm{\tilde{t}}_{q} \equiv \bm{\tilde{W}}_q \bm{\tilde{y}}$ and 
$\bm{\tilde{t}}_{hq} \equiv \bm{\tilde{W}}_{q} \bm{h}$.
 As a result, decomposing the data with PCA and keeping only the lowest variance components is equivalent to restricting the likelihood to its most significant terms (i.e., the terms that are the most relevant for GW parameters inference). The data-driven PCA process outlined here is similar to the TDI analysis, which has already been shown to be equivalent to an approximation of the likelihood~\cite{romano_principal_2006, vallisneri_tdi-infinity_2020, baghi_statistical_2021}. 
Depending on the choice of $q$, Eq.~\eqref{eq:loglikelihood_approx_t} can be considered as a family of approximations for Eq.~\eqref{eq:likelihood}. We have argued that we expect that small $q$ will provide a sufficient approximation, but the result become exact in the limit that $q \rightarrow 6p$.

\subsection{\label{sec:orthogonalization}Orthogonalization of aPCI streams with respect to secondary noises}

To further simplify the computation of the log-likelihood in Eq~\eqref{eq:loglikelihood_approx}, it is more convenient to  manipulate streams which are orthogonal with respect to the noise, including its non-laser part. Therefore, we need a process equivalent to the generation of A, E, T combinations~\cite{prince_lisa_2002} in TDI. We already have generalized this operation in a previous work~\cite{baghi_statistical_2021}, where we simply compute the eigenvectors of the covariance matrix:
\begin{equation}
	\label{eq:pca_channels_covariance_decomposition}
	\bm{\tilde{\Sigma}}_{\bm{t}}(f) = \bm{\Phi}(f) \bm{\Lambda}(f)\bm{\Phi}(f)^{\dagger},
\end{equation}
where $\bm{\Phi}(f)$ is the matrix of eigenvectors, and $\bm{\Lambda}(f)$ is the diagonal matrix of eigenvalues. Note that we perform this decomposition for each frequency. Ultimately, the orthogonal PCA combinations are given by
\begin{equation}
	\label{eq:pca_orthogonal_combinations}
	\bm{\tilde{t}}_{\perp}(f) =  \bm{\Phi}(f)^{\dagger} \bm{\tilde{t}}(f),
\end{equation}
and their PSDs are given by the diaogonal elements of $\bm{\Lambda}(f)$ . We have now a process to compute the noise levels necessary to derive the gravitational strain sensitivities.

\section{\label{sec:evaluation}Evaluation of aPCI performance}

The right quantity to assess the performance of the PCA process to analyse gravitational-wave measurements is the sensitivity of the PC projections, which we derive in this section.

\subsection{\label{sec:orbit}Orbit assumptions}
We assume that the LISA spacecraft follows Keplerian orbits approximated to first order in eccentricity, with $e = 4.8237 \times 10^{-3}$. We assume that the spacecraft form an equilateral triangle whose center follows a circular orbit around the Sun, such that its position at any time $t$ is given by
\begin{align}
    \bm{r}_0 = \left(R\cos \phi_T,\, R \sin \phi_T ,\,  0\right)^{\intercal},
\end{align}
where $R = 1\, \mathrm{au}$ is the distance between the Sun and the center of the constellation, and $\phi_T(t)$ is the flight path angle. 

\subsection{\label{sec:response}Derivation of gravitational-wave responses}

Here we derive the aPCI variables response to gravitational-wave signals, which is needed to compute their sensitivities. At the time of reception $t$, the response of gravitational waves in a single link $ij$ is given in fractional frequency by~\cite{estabrook_response_1975}:
\begin{equation}
	\label{eq:arm_esponse}
	y^{\mathrm{GW}}_{ij}(t) = \frac{H_{ij}\left(t_j - \frac{\bm{k} \cdot \bm{r}_{j}(t_j)}{c}\right) - H_{ij}\left(t - \frac{\bm{k} \cdot \bm{r}_{i}(t)}{c}\right)}{2\left(1 - \bm{k}\cdot \bm{\hat{n}}_{ij}\right)},
\end{equation}
where we labeled the GW projection onto arm $ij$ as $H_{ij}$, the spacecraft $i$ position vector as $\bm{r}_{i}$, the unit vector along arm $ij$ as $\bm{\hat{n}}_{ij} = \left(\bm{r}_{i} - \bm{r}_{j}\right) / \lVert \bm{r}_{i}- \bm{r}_{j} \rVert $ and the time of emission as $t_j = t - L_{ij}/c$, \hl{which depends on $L_{ij}$, the effective distance between the sending spacecraft $j$ and the receiving spacecraft $i$.} We compute the vectors $\bm{r}_{i}$ from the orbital motion postulated in Section~\ref{sec:orbit}. In this equation, we also defined $\bm{k}$ as the direction of GW propagation, which can be expressed as a function of the source's ecliptic latitude~$\beta$ and longitude~$\lambda$:
\begin{align}
	\bm{k} = - \left(\cos \lambda \cos \beta, \, \sin \lambda \cos \beta, \, \sin\beta \right)^{\intercal}.
\end{align}

The GW projection can be written as a function of antenna pattern functions $F_{+}$ and $F_{\times}$ such that
\begin{equation}
	\label{eq:contracted_metric2}
	H_{ij}(\tau) = h_{+}(\tau) F_{+}\left(  \psi , \bm{k}, \bm{\hat{n}}_{ij}\right) + h_{\times}(\tau) F_{\times}\left(  \psi , \bm{k}, \bm{\hat{n}}_{ij}\right),
\end{equation}
where $h_{\alpha}$ is the strain deformation in the source frame for the polarization mode $\alpha$. 
The antenna pattern functions depend on the GW polarization angle $\psi$, the wave propagation vector $\bm{k}$ and the link vector $\bm{\hat{n}}$ as
\begin{align}
	\label{eq:modulation_factors_def}
	F_{+}\left(  \psi , \bm{k}, \bm{\hat{n}}\right)  & =   \cos(2\psi)  \xi_{+}\left(\bm{k}, \bm{\hat{n}}\right)  -  \sin(2\psi) \xi_{\times}\left(\bm{k}, \bm{\hat{n}}\right) ;  \nonumber \\
	F_{\times}\left(  \psi , \bm{k}, \bm{\hat{n}}\right) & = \sin(2\psi) \xi_{+} \left(\bm{k}, \bm{\hat{n}}\right) + \cos(2\psi)\xi_{\times}\left(\bm{k}, \bm{\hat{n}}\right),
\end{align}
where  $\xi_{+}$ and $\xi_{\times}$ are modulation functions defined as
\begin{eqnarray}
	\xi_{+} \left(\bm{k}, \bm{\hat{n}}\right) &=& \left( \boldsymbol{u} \cdot \bm{\hat{n}} \right)^2 - \left( \boldsymbol{v} \cdot \bm{\hat{n}} \right)^2 ; \nonumber \\
	\label{eq:xi_def}
	\xi_{\times}\left(\bm{k}, \bm{\hat{n}}\right)  &=& 2 \left( \boldsymbol{u} \cdot \bm{\hat{n}} \right) \left( \boldsymbol{v} \cdot \bm{\hat{n}} \right).
\end{eqnarray}
The vectors $\bm{u}$ and $\bm{v}$ in the above equation form the source localization basis $\left(\bm{u}, \, \bm{v}, \, \bm{k} \right)$, with
\begin{align}
\bm{u} & = \left( \sin \lambda , \, - \cos \beta, \, 0 \right)^{\intercal} ; \nonumber \\
\bm{v} & = \left( - \sin \beta \cos \lambda , \, - \sin \beta \sin \lambda, \cos\beta \right)^{\intercal}.
\end{align}

To write the response as a function of the source frequency, it will be useful to express the GW strain in terms of its Fourier transform $\tilde{h}_{\alpha}(f)$:
\begin{equation}
	\label{eq:strain_fourier}
	h_{\alpha}(t) = \int_{-\infty}^{+\infty}  \tilde{h}_{\alpha}(f') e^{2i \pi f' t} df'.
\end{equation}
We plug Eq.~\eqref{eq:strain_fourier} into Eq.~\eqref{eq:arm_esponse} to get
\begin{align}
	\label{eq:arm_esponse_frequency}
	y^{\mathrm{GW}}_{ij}(t) & = \sum_{\alpha=+,\times} \int_{-\infty}^{+\infty}  \tilde{h}_{\alpha}(f') e^{2i \pi f' t} df' \frac{  F_{\alpha}\left(  \psi , \bm{k}, \bm{\hat{n}}_{ij}\right)}{2\left(1 - \bm{k}\cdot \bm{\hat{n}}_{ij}\right))}  \nonumber \\
	& \times \left[ e^{- 2i \pi f' \left(L_{ij} + \bm{k} \cdot \bm{r}_{j}(t_j)\right) / c}- e^{- 2i \pi f' \bm{k} \cdot \bm{r}_{i}(t)/ c} \right].
\end{align}

In the following, we assume that the gravitational wave is monochromatic, with frequency $f$. 
Choosing a null phase, the Fourier transform of $h_{\alpha}(t)$ can then be written as $\tilde{h}_{\alpha}(f') = \frac{1}{2} A_{\alpha} \delta \left(f'-f\right)$ for positive frequencies. Inserting this into Eq.~\eqref{eq:arm_esponse_frequency_harmonic} gives
\begin{align}
	\label{eq:arm_esponse_frequency_harmonic}
	y^{\mathrm{GW}}_{ij}(t, f) & = \sum_{\alpha=+,\times} \frac{1}{2} A_{\alpha} e^{2i \pi f t} \frac{ F_{\alpha}\left(  \psi , \bm{k}, \bm{\hat{n}}_{ij}\right)}{2\left(1 - \bm{k}\cdot \bm{\hat{n}}_{ij}\right))}  \nonumber \\
	& \times \left[ e^{- 2i \pi f \left(L_{ij} + \bm{k} \cdot \bm{r}_{j}(t_j)\right) / c}- e^{- 2i \pi f \bm{k} \cdot \bm{r}_{i}(t)/ c} \right].
\end{align}
The modes amplitudes $A_{\alpha}$ can be expressed as a function of the inclination $\iota$ of the system with respect to the line of sight:
\begin{align}
\label{eq:source_amplitudes}
	A_{+} & = A \left(1 + \cos^2 \iota\right) ; \nonumber \\
	A_{\times} & = - 2 A \cos \iota,
\end{align}
where $A$ is the amplitude of the source, that we consider constant in our monochromatic assumption.

We can now derive the response of PCA projections to GWs using the time-domain equivalent of  Eqs.~\eqref{eq:pca_transformation_fourier}:
\begin{equation}
	\label{eq:pca_response}
	\bm{t}^{\mathrm{GW}}(f) = \bm{V}^{\dagger} \begin{pmatrix} 
	\bm{y}^{\mathrm{GW}}(t+n_h \tau_s , f) \\ 
	\vdots \\
	\bm{y}^{\mathrm{GW}}(t-n_h \tau_s,  f)\end{pmatrix}^{\dagger},
\end{equation}
where $\bm{y}^{\mathrm{GW}}(t, f)$ is simply the vector of single-link GW responses
\begin{align}
	\label{eq:y_vector_gw}
	\bm{y}^{\mathrm{GW}}(t, f) \equiv \begin{pmatrix} 
		y^{\mathrm{GW}}_{12}(t, f) \\
		y^{\mathrm{GW}}_{23}(t, f) \\
		 y^{\mathrm{GW}}_{31}(t, f) \\
		 y^{\mathrm{GW}}_{13}(t, f) \\
		 y^{\mathrm{GW}}_{21}(t, f) \\
		 y^{\mathrm{GW}}_{32}(t, f)
		\end{pmatrix},
\end{align}
whose entries are given by Eq.~\eqref{eq:arm_esponse_frequency_harmonic}.

\subsection{\label{sec:sensitivity}Computation of sky-averaged sensitivity}

In Sec.~\ref{sec:covariance} we derived the frequency-domain covariance of the aPCI combinations as a function of the single-link data covariance $\bm{\tilde{\Sigma}}_{\bm{y}}$, while in Sec.~\ref{sec:response} we derived their response to gravitational waves. 
Following Babak \textit{et al.}~\cite{babak_lisa_2021}, we define the instantaneous strain sensitivity of a given component $m$ as its noise PSD divided by its response averaged over all sky locations and orientations:
\begin{equation}
	\label{eq:single_component_sensitivity}
	S_{h, m}(f) = A^2\frac{S_{\perp, m}(f)}{\langle \lvert t^{\mathrm{GW}}_{\perp, m}(f)\rvert^2 \rangle},
\end{equation}
where $\bm{t}^{\mathrm{GW}}_{\perp} \equiv \bm{\Phi}(f)^{\dagger} \bm{t}^{\mathrm{GW}}$ is the GW response in the orthogonalized aPCI combinations, $S_{\perp, m}(f) \equiv \Lambda_{mm}(f)$ is the noise PSD given by the diagonal elements of matrix $\Lambda(f)$ in Eq.~(\ref{eq:pca_channels_covariance_decomposition}), and the brackets denote the averaging over sky location angles, polarization and inclination:
\begin{align}
	\label{eq:averaged_response}
	\langle \lvert t^{\mathrm{GW}}_{\perp, m}\rvert^2 \rangle= \frac{1}{4\pi}\int_{\bm{k}} \frac{1}{2\pi} \int_{0}^{2\pi} \frac{1}{2} \int_{-1}^{+1} \lvert t^{\mathrm{GW}}_{\perp, m}\rvert^2 d^{3}\bm{k} d\psi d\cos\iota.
\end{align}
In this definition, the sensitivity corresponds to the level of strain noise within an infinitesimal frequency band $df$ (as stated in, e.g., ~\cite{vallisneri_non-sky-averaged_2012}). Note that we include the factor $A^2$ in front of Eq.~(\ref{eq:single_component_sensitivity}) to cancel the one present in the denominator via Eqs.~\eqref{eq:source_amplitudes}, making the sensitivity independent of the amplitude, as it precisely determines the variance of $A$.
We can sum the sensitivity over a subset of components between $m_{\mathrm{min}}$ and $m_{\mathrm{max}}$:
\begin{equation}
	\label{eq:all_components_sensitivity}
	S_{h}(f) = \left[ \sum_{m=m_{\mathrm{min}}}^{m_{\mathrm{max}}} S^{-1}_{h, m}(f) \right]^{-1},
\end{equation}
where the terms in the sum are the inverse sensitivities of each component, which can be viewed as their SNR per unit time and unit amplitude. In the following, we compute this sensitivity on a given example and compare it with its TDI analog.

To evaluate the sensitivity in Eq.~\eqref{eq:single_component_sensitivity}, we need an analytical model for the PSD of the single-link data, at least under some assumptions. This will then allow us to derive an expression for the PSD of the PCA combinations $S_{\perp, m}(f)$.
We assume that LISA's single-link measurements include two main sources of noise: a large term due to laser frequency fluctuations and a much smaller term due to other instrumental noises.  We assume that the signals measured by the two optical benches on-board spacecraft 1 are given by the equations
\begin{eqnarray}
	\label{eq:phasemeneter_meas}
	s_{12}(t) &=& p_{21}(t - c^{-1}L_{12}) - p_{12}(t) + n_{12}(t) \nonumber \\
	s_{13}(t) &=& p_{31}(t - c^{-1}L_{13}) - p_{13}(t) + n_{13}(t),
\end{eqnarray}
where $p_{ij}$ are laser noises, $n_{ij}$ are secondary noises, and $c$ is the speed of light, labeling by the convention noted in Section~\ref{sec:data_matrix}. We obtain the other measurements on-board spacecraft 2 and 3 by permuting the indices (modulo 3) in Eq.~(\ref{eq:phasemeneter_meas}).
To simplify the analysis, we assume that the two lasers onboard each spacecraft $i$ are identical, so that $p_{ij} = p_{ik}$. This amounts to assuming that the $\bm{s}_{ij}$ are equal to the vectors of intermediate variables $\bm{\eta}_{ij}$ in the TDI pre-processing (see, e.g., Ref.~\cite{otto_tdi_2012}). In this model, the measured laser noises can be seen as the result of the mixing (and delaying) of a vector of laser noise sources $\bm{p} = \left(\bm{p}_{12}^{\intercal}, \, \bm{p}_{23}^{\intercal}, \, \bm{p}_{31}^{\intercal} \right)^{\intercal}$. We consider the $6N$ column-vector version of the $N \times 6$ matrix $\bm{Y}$ in Eq.~\eqref{eq:y_matrix}, which stacks all the measurements as 
\begin{equation}
    \label{eq:y_vector}
    \bm{y} \equiv \left(\bm{s}_{12}^{\intercal}, \, \bm{s}_{23}^{\intercal}, \, \bm{s}_{31}^{\intercal}, \, \bm{s}_{13}^{\intercal}, \, \bm{s}_{21}^{\intercal}, \, \bm{s}_{32}^{\intercal} \right)^{\intercal}.
\end{equation}
This mixing is represented by a $6N \times 3N$ design matrix $\bm{M}$, such that the model for $\bm{y}$ is
\begin{equation}
\label{eq:phasemeter_meas_matrix}
	\bm{y} = \bm{M} \bm{p} + \bm{n},
\end{equation}
where we set $\bm{n} \equiv \left(\bm{n}_{12}^{\intercal}, \, \hdots, \, \bm{n}_{32}^{\intercal} \right)^{\intercal}$, using the same ordering as in Eq.~\eqref{eq:y_vector}. Using this assumption, we have
\begin{equation}
	\label{eq:delay_op_matrix}
	\bm{M} = \begin{pmatrix}
		-\bm{1}            &  \bm{D}_{12} & \bm{0}\\
		\bm{0}             & -\bm{1}& \bm{D}_{23} \\
		\bm{D}_{31}   & \bm{0}        & -\bm{1} \\
		-\bm{1}           & \bm{0}        & \bm{D}_{13}  \\
		\bm{D}_{21}   & -\bm{1}      &  \bm{0}  \\
		\bm{0}   & \bm{D}_{32}         & -\bm{1}
	\end{pmatrix},
\end{equation}
where $\bm{D}_{lm}$ represents the fractional delay operator acting on any discretized signal $x(t_n)$ as $\bm{D}_{lm}\bm{x}|_{n} \approx x\left(t - L_{lm}/c\right)$.

In the following, we assume that $\bm{p}_{ij}$ and $\bm{n}_{ij}$ are zero-mean stationary Gaussian noises vectors of size $N$ described by PSDs $S_{p, ij}(f)$ and $S_{n, ij}(f)$, respectively. We further assume that these entries are uncorrelated. Since the laser noise level is much larger than the other noises, we have $S_{p, ij}(f) \gg S_{n, ij} \, \forall i, j$. For this reason, we will refer to $\bm{n}$ as the vector of secondary noises.

\subsection{\label{sec:sensitivity_steps}Practical implementation}

For a number of aPCI variables, we compute the orientation and sky-averaged sensitivity following Eqs.~\eqref{eq:all_components_sensitivity}. We perform the integration with respect to the polarization angle analytically, which is equivalent to setting $\psi_{\mathrm{eq}} = \pi / 4$ (as it corresponds to equal contributions between the $+$ and $\times$ modes). We also use an analytical result for the integration over inclination, as it amounts to multiplying the squared response by $4/5$~\cite{robson_construction_2019}. Finally, we average over sky locations numerically by randomly drawing 3000 values of the source ecliptic latitude $\beta$ and longitude $\lambda$ from a uniform distribution in the interval $\left[0, \, \pi \right] \times \left[0,\, 2\pi\right]$.

Suppose we have computed the SVD decomposition. Among the $6p$ PCs provided by the algorithm, let us consider only $q$ of them. For each sky location $(\lambda,\, \beta)$ and each GW frequency $f$, we compute the response by following 3 steps:
\begin{enumerate}
    \item Compute the arm responses $y^{\mathrm{GW}}_{ij}$ using Eq.~\eqref{eq:arm_esponse_frequency_harmonic};
    \item Form the $q$ PCA projections $\bm{t}^{\mathrm{GW}}_{m}$ with Eq.~\eqref{eq:pca_response};
    \item Orthogonalize the projections using Eq.~\eqref{eq:pca_orthogonal_combinations};
\end{enumerate}

To get a baseline for comparison, we also compute the sensitivity of first-generation TDI, as defined in, e.g., Ref.~\cite{tinto_time_2004}. 
The computation steps follow a similar scheme as above. The first step is the same: we calculate the arm response to harmonic GWs exactly as before. Step~2 is replaced by the computation of the Michelson TDI combinations $X$, $Y$ and $Z$. This is done by applying the TDI 1.5 transfer function to the single-link response vector $\bm{t}^{\mathrm{GW}}_{\mathrm{TDI}} = \bm{V}_{\mathrm{TDI}}^{\dagger} \bm{y}^{\mathrm{GW}}$, where $\bm{V}_{\mathrm{TDI}}$ can be written as
\begin{widetext}
\begin{align}
\bm{V}_{\mathrm{TDI}} & \equiv 
\begin{pmatrix}
D_{13}D_{31} - 1 & \left( 1 - D_{23}D_{32} \right)  D_{21} & 0 &\\ 
0 & D_{21}D_{12} - 1 & \left( 1 - D_{31}D_{13} \right) D_{32}\\
\left( 1 - D_{12}D_{21} \right) D_{13} & 0 & D_{32} D_{23}  - 1\\
1 -  D_{12} D_{21} & 0 & \left( D_{32} D_{23}  - 1 \right) D_{31} \\
\left(D_{13}D_{31}  - 1 \right) D_{12} & 1 - D_{23} D_{32} & 0 \\
0 & \left( D_{21}D_{12} - 1 \right) D_{23} & 1 - D_{31} D_{13}\\
\end{pmatrix}.
\end{align}
\end{widetext}
Then, Step 3 is replaced by forming the $A$, $E$ and $T$ combinations from $X$, $Y$ and $Z$, as defined in Ref.~\cite{tinto_time-delay_2014}. Finally, we compute the TDI sensitivity over all channels with a calculation similar to Eq.~(\ref{eq:all_components_sensitivity}).

\section{\label{sec:simulations}Numerical simulations and results}

In this section, we verify that the aPCI algorithm yields a significant noise subtraction. Then we check the PSD model derived in Section~\ref{sec:covariance} against realizations of the noise. Finally, we compute averaged sensitivities to gravitational waves, and compare them with Sagnac TDI observables A, E and T.

\subsection{\label{sec:noise_subtraction}Laser noise subtraction}

We test the aPCI algorithm on simulated LISA data streams. We adopt the same assumptions as in Ref.~\cite{baghi_statistical_2021}. To generate the noises, we assume that the laser sources of two optical benches belonging to the same spacecraft are the same, and that the light travel time delays between spacecraft are constant and equal to $L/c$, where $L = 2.5 \times 10^9$ m is the mean inter-spacecraft distance. 

We simulate the science measurement time series $s_{ij}$ from Eq.~\eqref{eq:phasemeneter_meas}. We generate noises from zero-mean Gaussian stationary processes in the frequency domain described by PSDs $S_p(f)$ and $S_n(f)$ given by:
\begin{align}
    S_{p}(f) &= \left(\frac{28.2 \, \mathrm{Hz.Hz^{-1/2}}}{\nu_0}\right)^2 ;\nonumber \\
    S_{n}(f) &= \frac{S_{a}(f)}{\left(2 \pi f c \right)^2} + \left(\frac{2\pi f}{c}\right)^2 S_{\mathrm{oms}}(f),
\end{align}
where $\nu_0 = c \cdot (1064 \times 10^{-9}\mathrm{m})^{-1}$ is the central frequency of the lasers, $S_{a}(f)$ is the test-mass acceleration noise PSD (in $\rm m^2 s^{-4} Hz^{-1}$) and $S_{\mathrm{oms}}(f)$ is the PSD representing all other pathlength noises in the measurement (in $\rm m^2 Hz^{-1}$). Their expressions are given in Appendix~\ref{sec:secondary_noises}, where we adopted assumptions made for the \textsc{LISANode} simulator~\cite{bayle:tel-03120731}.
Likewise, we apply light travel time delays in the frequency domain. To prevent any artificial periodicity, we generate $2N$ samples corresponding to Fourier frequencies $\pm k f_s / (2N)$ before transforming the generated noise to the time domain, and discarding the second half of the data to get a time series of size $N$.

We choose the duration of the time series to be 12 hours, with a sampling rate of $f_s = 2$ Hz, which allows us to cover the frequency band between 0.1 mHz and 1 Hz. We choose the half-stencil size $n_h$ so that the stencil spans the duration of one complete turn around the constellation $n_h = \lfloor 3 L f_s/ c \rfloor = 25$. Note that we did not find significant changes in the results presented in the following when choosing larger values for $N$ and $n_h$. A full assessment of the impact of data and stencil sizes is left for future work. We form the data matrix following Eq.~\eqref{eq:data_matrix} and compute its PCA as in Eq.~\eqref{eq:pca_transformation} with the algorithm provided by the \textsc{Python} package \textsc{Scikit-learn}~\cite{pedregosa_scikit-learn_2011}. In this example, the decomposition yields 306 PCA components.

\begin{figure}[!ht]
\includegraphics[width=\columnwidth, trim={0.5cm, 0.5cm, 0.5cm, 0.3cm}, clip]{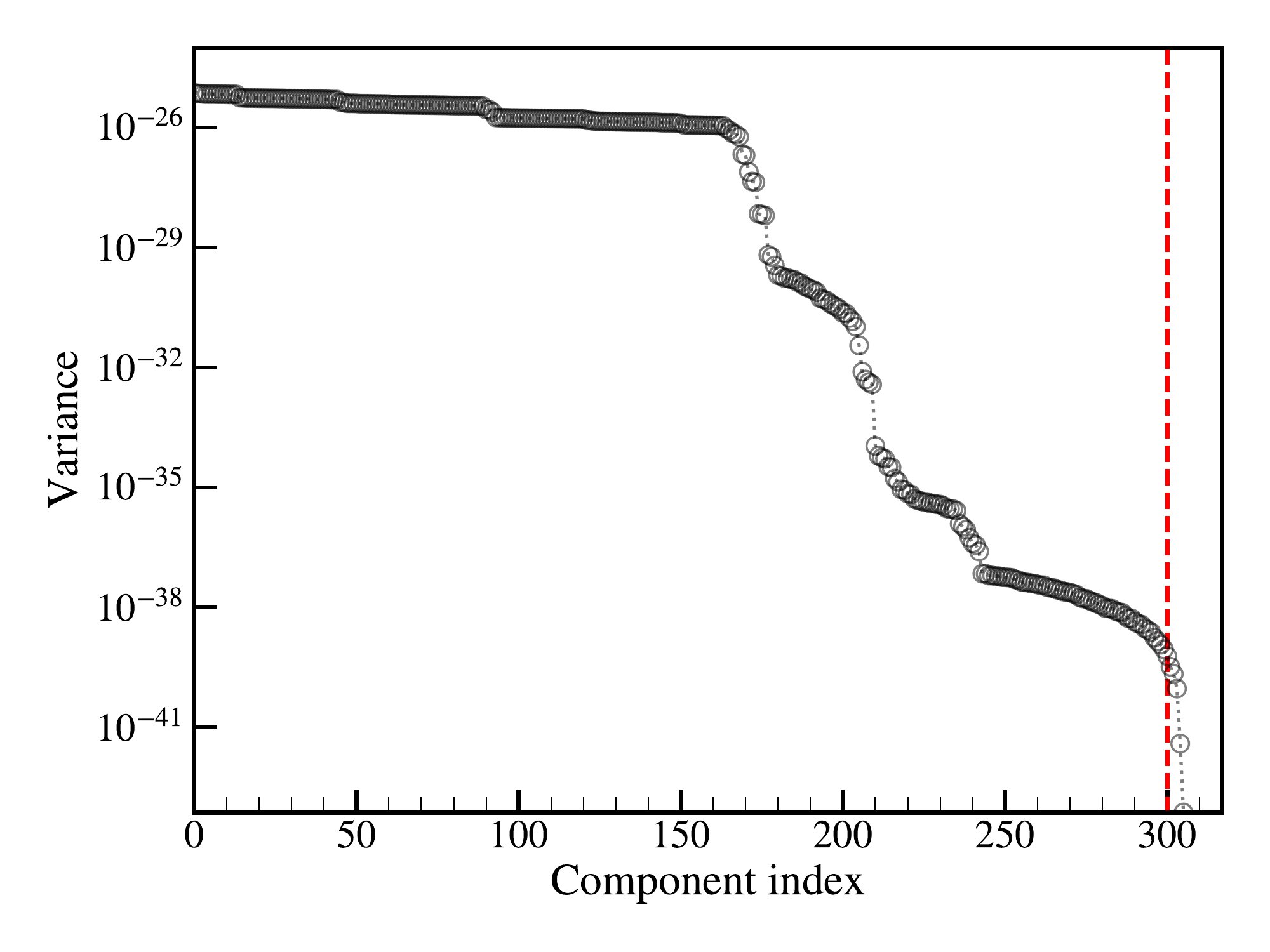}%
\caption{\label{fig:explained_variance_nh25}Variance associated to PCA components in decreasing order. Instead of selecting the largest variance components like in classic PCA, here we are interested in the lower end of the plot, i.e. the less noisy components at the right-hand side of the red dashed line.}
\end{figure}

We can analyze the relative amount of noise present in each component $m$ by calculating the variance quantity $S^2_{m, m} / (N-1)$, where $\bm{S}$ is the singular value matrix in Eq.~\eqref{eq:svd}. We report these values for all $m$ on Fig.~(\ref{fig:explained_variance_nh25}). We see that the variance decreases with increasing values of $m$, with a drop spanning 15 orders of magnitudes. This plot is a good hint about the ability of the algorithm to separate large noise from low noise components.

\begin{figure}[!ht]
\includegraphics[width=\columnwidth, trim={0.5cm, 0.5cm, 0.5cm, 0.3cm}, clip]{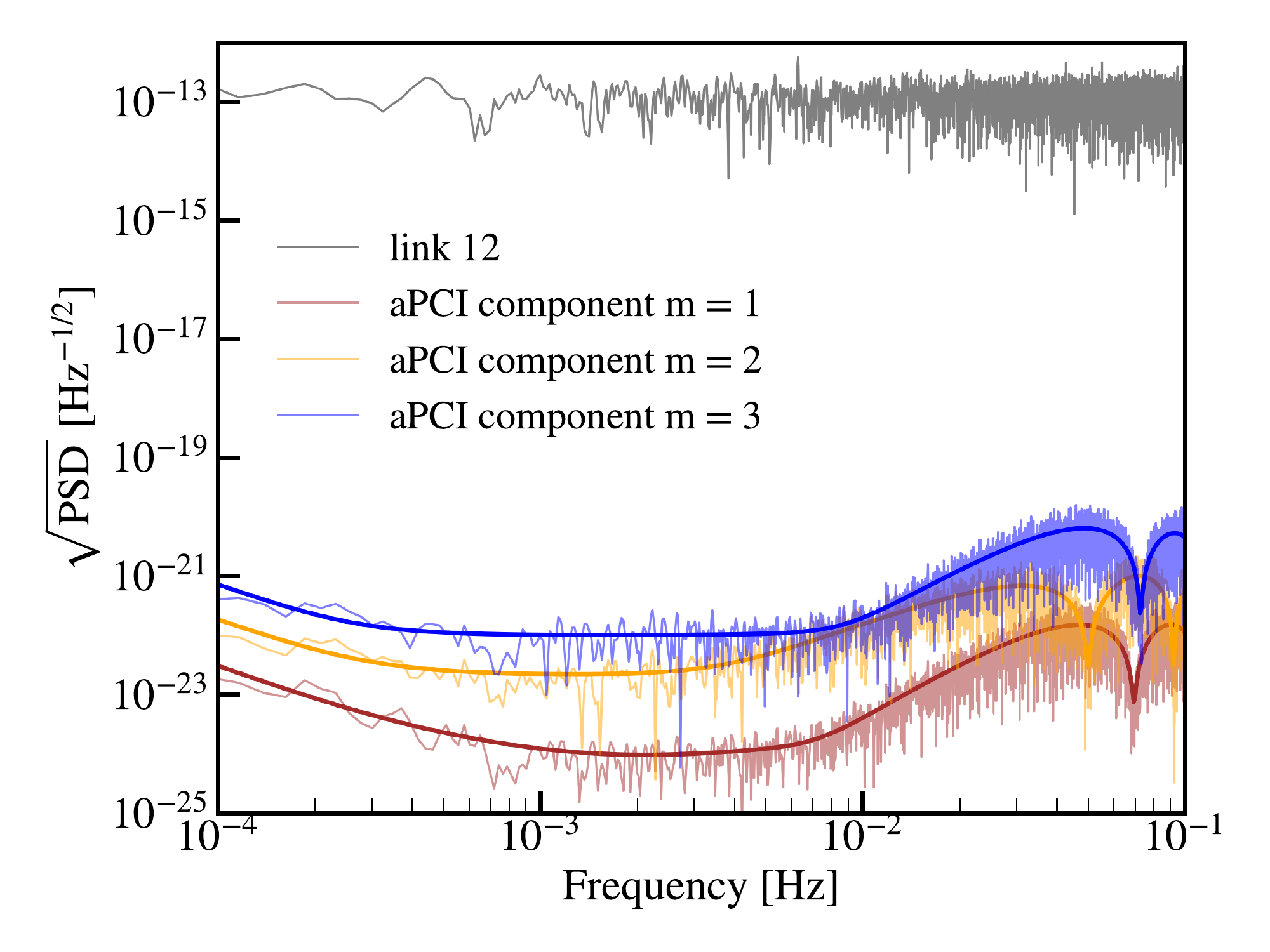}%
\caption{\label{fig:projections_nh25}Periodograms (light colors) of the three least noisy PCA projections as a function of frequency. We also plot the theoretical PSDs (dark colors) obtained with Eq.~\eqref{eq:pca_channels_covariance_fourier}, expressed in units of relative frequency deviation. To compare with the input noise, we plot in gray the periodogram of the raw measurement from laser link 12, which is overwhelmed by laser noise.}
\end{figure}

Now let us order the aPCI variables from the lowest variance to the largest variance, so that index $m = 1$ corresponds to the rightmost end of Fig.~\ref{fig:explained_variance_nh25}, and $m = 6p$ corresponds to the leftmost end. We project the data vector $\bm{y}$ onto the last 3 singular vectors using Eq.~\eqref{eq:pca_transformation}, obtaining 3 data projections whose periodograms are plotted in Fig.~\ref{fig:projections_nh25} in red for $m = 1$, in orange for $m = 2$, in blue for $m = 3$. We define the cross-periodogram of two time series $x$ and $y$ of size $N$ as 
\begin{align}
    P_{xy}(f) = \tilde{x}(f) \tilde{y}^{\ast}(f),
\end{align}
where $\tilde{x}$ refers to the windowed discrete Fourier transform of $x$ normalized as
\begin{align}
    \tilde{x}(f) \equiv \sqrt{\frac{2 \tau_s}{N}}\sum_{n=0}^{N-1} w_n x_n e^{-2 i \pi f n \tau_s}.
\end{align}

As a comparison, we plot the periodogram of link 12 (gray curve). The difference between the gray and colored curves shows that the noise is significantly suppressed by several orders of magnitude. In addition, we verify that the noise model derived in Eq.~\eqref{eq:pca_channels_covariance_fourier} accurately predicts the noise levels in the periodograms. To do so, we include the added contributions of laser noise and secondary noises to the aPCI residuals. 

\begin{figure}[!ht]
\includegraphics[width=\columnwidth, trim={0.5cm, 0.7cm, 0.5cm, 0.3cm}, clip]{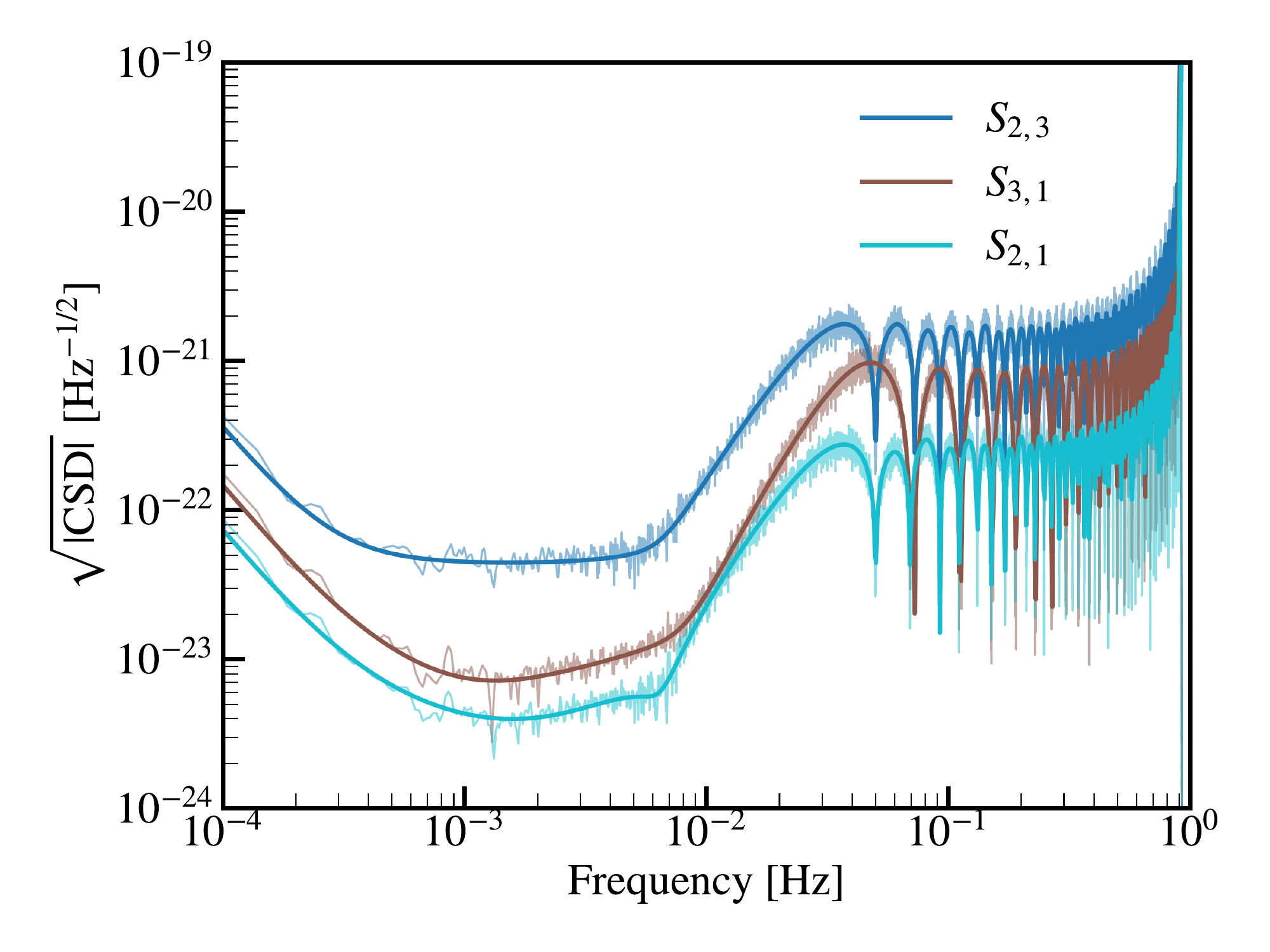}%
\caption{\label{fig:csd_nh25}Averaged cross-periodograms of the 3 aPCI variables of lowest variance (thin light lines), along with the corresponding analytical CSD model (darker solid lines). Correlations among components are comparable to their individual PSDs (plotted in Fig.~\ref{fig:projections_nh25}).}
\end{figure}

Besides, we analyze the correlations among the aPCI variables by comparing the cross-periodograms $P_{{t_m}{t_l}}(f)$ with the cross spectral density (CSD) model provided by the non-diagonal terms of the covariance in Eq.~\eqref{eq:pca_channels_covariance_fourier}. 
As an example, we plot in Fig.~\ref{fig:csd_nh25} the CSD of the pairs $(m, l) = (2, 3)$ in blue, $(3, 1)$ in brown, $(2, 1)$ in turquoise. They are all consistent with the cross-periodogram (lighter colors) averaged over 25 realizations of the noise. We see that the aPCI variables are strongly correlated, with CSD levels comparable to the PSDs. 
There is no reason why this should not be the case, as the PCA decomposition is built mainly from the dominant laser noise statistics. In addition, the matrix $\bm{X}$ we decompose is formed from the data itself, not from the actual covariance of the data $\bm{\Sigma}_{y}$. As a result, the process does not give exactly orthogonal data streams by construction. That is why an extra diagonalization step can be applied, as described in Section~\ref{sec:orthogonalization}, to account for secondary noises.

While the noise levels in the aPCI streams $\bm{t}_{m}$ give us an indication about the amount of laser noise that is suppressed, there is little sense in comparing them directly to the TDI PSDs. It is rather preferable to compute GW sensitivities, which we do in the next section.

\subsection{Comparison of aPCI and TDI sensitivities}

In this section, we arrange the aPCI variables in increasing order of variance, i.e. we start from the lower end of the curve in Fig.~\ref{fig:explained_variance_nh25}. We define the sensitivity of the $q$ variables of lowest variance as the sensitivity obtained by accumulating the SNRs of their orthogonal transformations, as outlined in the steps of Section~\ref{sec:sensitivity_steps}.

\begin{figure}[!ht]
\includegraphics[width=\columnwidth, trim={0.5cm, 0.3cm, 0.5cm, 0.3cm}, clip]{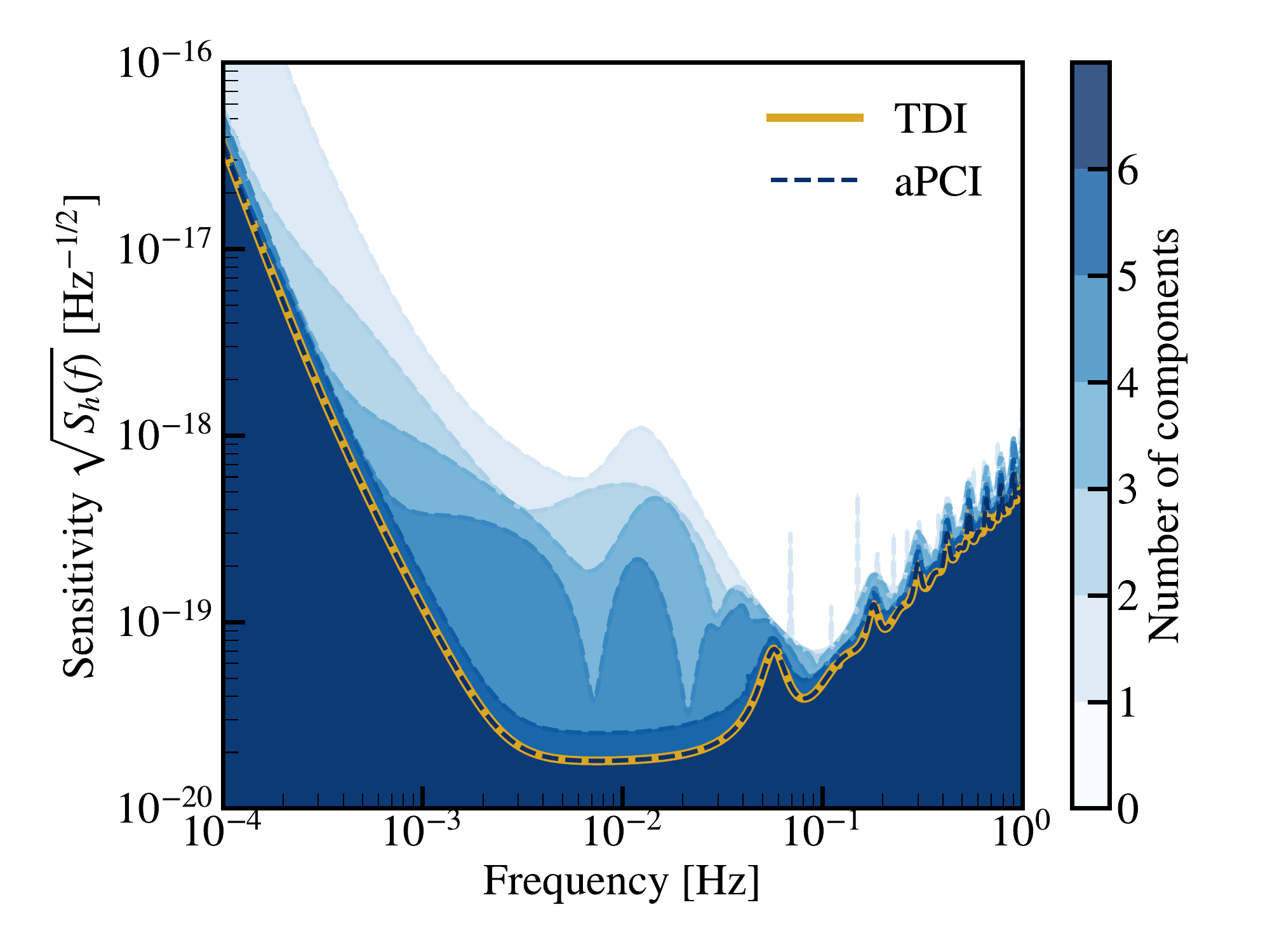}%
\caption{\label{fig:sensitivity} Comparison of aPCI (blue dashed lines) and TDI (thick orange line) sensitivities summed over combinations. The lightest, uppermost blue dashed line corresponds to the sensitivity obtained with a single aPCI component, chosen to be the one with the smallest singular value. The darker curve just below corresponds to the accumulated sensitivity obtained by including an additional component (corresponding to the next smallest singular value). Darker blue dashed curves beneath are constructed similarly, by adding more components. The lowermost dashed blue curve is the sensitivity reached with 6 components, which almost perfectly fits the orange line representing the accumulated AET TDI sensitivity.}
\end{figure}

We plot the sensitivities for a number of components ranging from $q = 1$ to $q = 6$ in Fig.~\ref{fig:sensitivity}. Each blue dashed line corresponds to a value of $q$, with darker colors for larger $q$. To better distinguish the levels, we fill the areas in between two levels with related shades of blue. For example, the uppermost, lightest curve is the sensitivity obtained with a single aPCI variable ($q=1$), i.e. the one with lowest variance. The curve that comes just below ($q=2$) is the sensitivity obtained by using the two variables of lowest variance, and so on. As expected, we can see that the larger the number of components we include, the better the sensitivity.

We also plot in Fig.~\ref{fig:sensitivity} the summed sensitivity of TDI channels A, E and T with a thick, solid orange line. Remarkably, the figure shows that we reach the same sensitivity as orthogonal TDI with 6 aPCI variables, as we can see from the superimposition of the lowest dashed blue line and the orange solid line. We have checked that adding more aPCI variables does not improve the sensitivity any further, as they are too much dominated by laser noise.

\subsection{\label{sec:effect_of_noise}SNR distribution and statistical behavior}

Fig.~\ref{fig:sensitivity} suggests that we need 6 aPCI variables to reach the exact same sensitivity as the 3 orthogonal TDI combinations. To analyse the extent to which this is true, we can assess the fraction of sensitivity, or rather equivalent squared SNR, contained in each component. We do so by defining the squared SNR as simply the inverse sensitivity integrated over the frequency band:
\begin{equation}
    \mathrm{\overline{SNR}}_{m}^{2} \equiv \int_{f_{\mathrm{min}}}^{f_{\mathrm{max}}}  S^{-1}_{h, m}(f)df,
\end{equation}
where $S_{h, m}(f)$ is given by Eq.~\eqref{eq:single_component_sensitivity}, $f_{\mathrm{min}} = 10^{-4}$ Hz and $f_{\mathrm{max}} = 1$ Hz. We are not using the term ``SNR" in the usual way here, as the inverse sensitivity is just a proxy for the SNR of a monochromatic source. It can be thought of as a SNR per unit time and unit strain. Then, the squared SNR fraction of component $m$ relative to TDI is simply $\mathrm{\overline{SNR}}_{m}^{2}$ divided by the total  SNR$^{2}$ achieved by variables A, E, T.

Besides, since the PCA is performed on noisy data, singular vectors are random variables with a variability depending on the particular realization of the noise present in $\bm{y}$. 
Therefore, to study how the sensitivity is distributed between different components, we need to consider several noise realizations and examine this distribution on average.

\begin{figure}[!ht]
\includegraphics[width=\columnwidth, trim={0.5cm, 0.5cm, 0.5cm, 0.3cm}, clip]{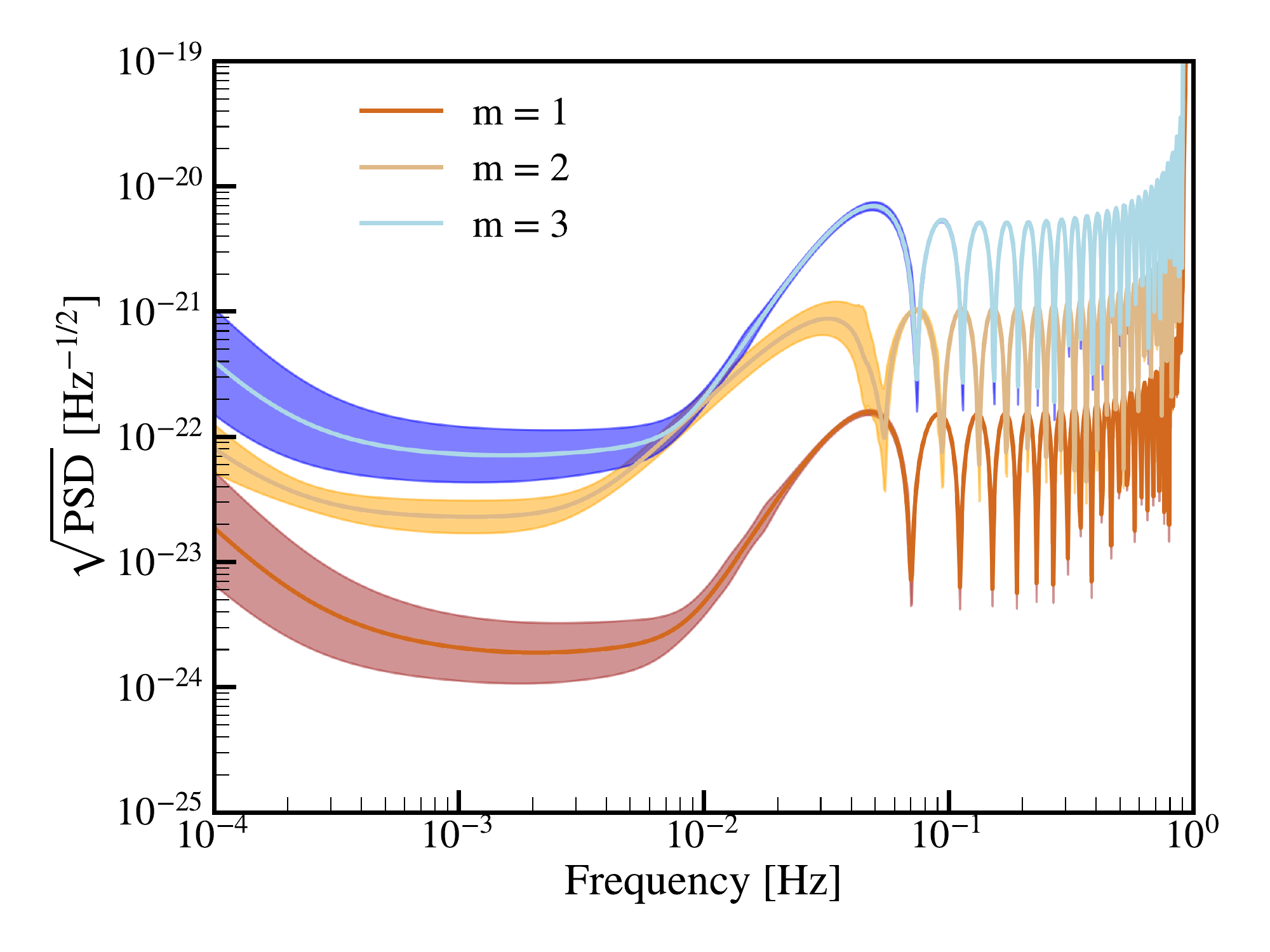}%
\caption{\label{fig:psd_distibution}Impact of noise realizations on the three least noisy aPCI variables noise PSDs. The solid lines correspond to the mean PSDs accross 100 noise realizations, while the colored areas around each line represent the 1-$\sigma$ region. The indexing $m$ is ordered along increasing variance, from $m=1$ (lowest variance) to $m=3$ (third lowest variance).}
\end{figure}

For this purpose, we generate 100 realizations of the 12-hour long vector $\bm{y}$, and compute the aPCI variables for each of them, along with their theoretical PSDs. We gather the results in Fig.~\ref{fig:psd_distibution}, where we plot the mean PSDs for the three variables with lowest singular values (the same as in Fig.~\ref{fig:projections_nh25}), along with their 68\% confidence intervals. This plot reveals the significant variability of the relative noise levels within each component.

\begin{figure}[!ht]
\includegraphics[width=\columnwidth, trim={0cm, 0.2cm, 0cm, 0cm}, clip]{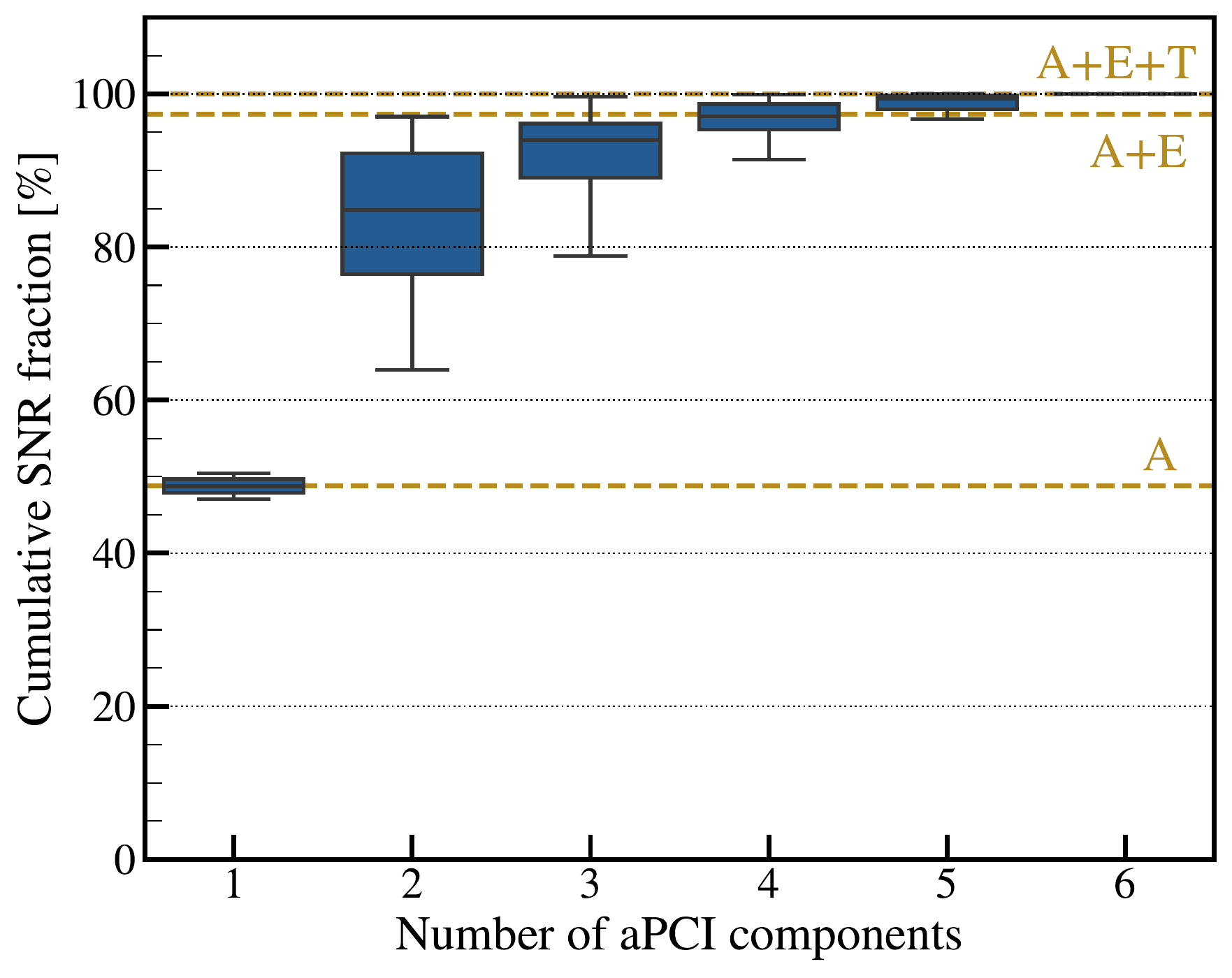}%
\caption{\label{fig:snr_fractions_distributions}
Distributions of cumulative fraction of SNR$^2$ (in percent) as a function of the number of variables, ordered in decreasing SNR$^2$ fractions. The blue boxes correspond to aPCI variables and are made of 3 line showing the 3 quartiles, while the vertical lines on both sides extend to the extreme values of the distribution. The yellow horizontal lines show the cumulative SNR$^2$ fraction of TDI variables A, E, T. The distributions are obtained from 100 realizations of the data vector $\bm{y}$.}
\end{figure}

For each noise realization, we compute the sensitivity of the 6 lowest variance aPCI variables, and the SNR$^2$ fraction (relative to TDI) carried by each variable. 
In Fig.~\ref{fig:snr_fractions_distributions}, we plot the distribution of cumulative SNR$^2$ fraction as a function of the number of components in the form of box plots (blue), using the \textsc{seaborn} package~\cite{Waskom2021}. We range the components from largest to smallest SNR$^2$. As a reference, we plot the TDI combinations' cumulative SNR$^2$ when adding A, E and T in this order (horizontal yellow dashed lines).

We see that in average, most of the SNR$^2$ (about 94\%) is concentrated in the 3 most sensitive aPCI variables (boxes on the left-hand side of the figure). While the first component always gather about 50\% of the SNR$^2$, the heights of the boxes indicate a significant spread of the SNR$^2$ fraction around the median when adding components 2 and 3. The 3 last components (4, 5, 6) have marginal SNR concentrations, since their total fraction is about 6\% on average. It should be noted that the last box is almost flat, which means that the total SNR (or sensitivity) hardly changes from one realization to another. \hl{It fluctuates around 99.99751(7)\% of the TDI total SNR.} Therefore, the SNR contained in the last 6 aPCI variables is virtually equal to the cumulative SNR of the orthogonal TDI variables. \hl{The model independence results in negligible SNR loss, however the gravitational wave signal is diluted across more than the traditional 3 TDI variables.}

\subsection{\label{sec:gw_impact}Impact of gravitational waves}

The PCA process outlined in this work acts directly on the measurements, hence the combinations that we derive depend on the content of the data. A natural question that arises is the impact of the presence of gravitational-wave signals onto the PC vectors, and thus on the overall sensitivity of the aPCI variables.  

The core property which distinguishes the laser noise as we isolate it for cancellation is just that it is loud. If a sufficiently large gravitational-wave signal is present, we should expect that our process will cancel that as well, but how loud does the signal have to be for such losses to be significant?  

Here, we take a first step in answering that question by simulating a quasi-monochromatic signal from a ultra-compact galactic binary, whose amplitude $A$ is allowed to vary. For the purpose of this assessment, we consider arbitrary and unphysical amplitude values chosen to give the source particular SNRs, ranging from 1 to $10^9$. We choose a fixed GW frequency equal to 1 mHz, as well as a fixed sky location. For each value of the amplitude, we create a data set by injecting the GW signal into the phasemeter measurements, in addition to a noise realization which remains the same for all sets. The duration of the time series is the same as in the previous sections, i.e. 12 hours.

\begin{figure}[!ht]
\includegraphics[width=\columnwidth, trim={0.5cm, 0.5cm, 0.5cm, 0.3cm}, clip]{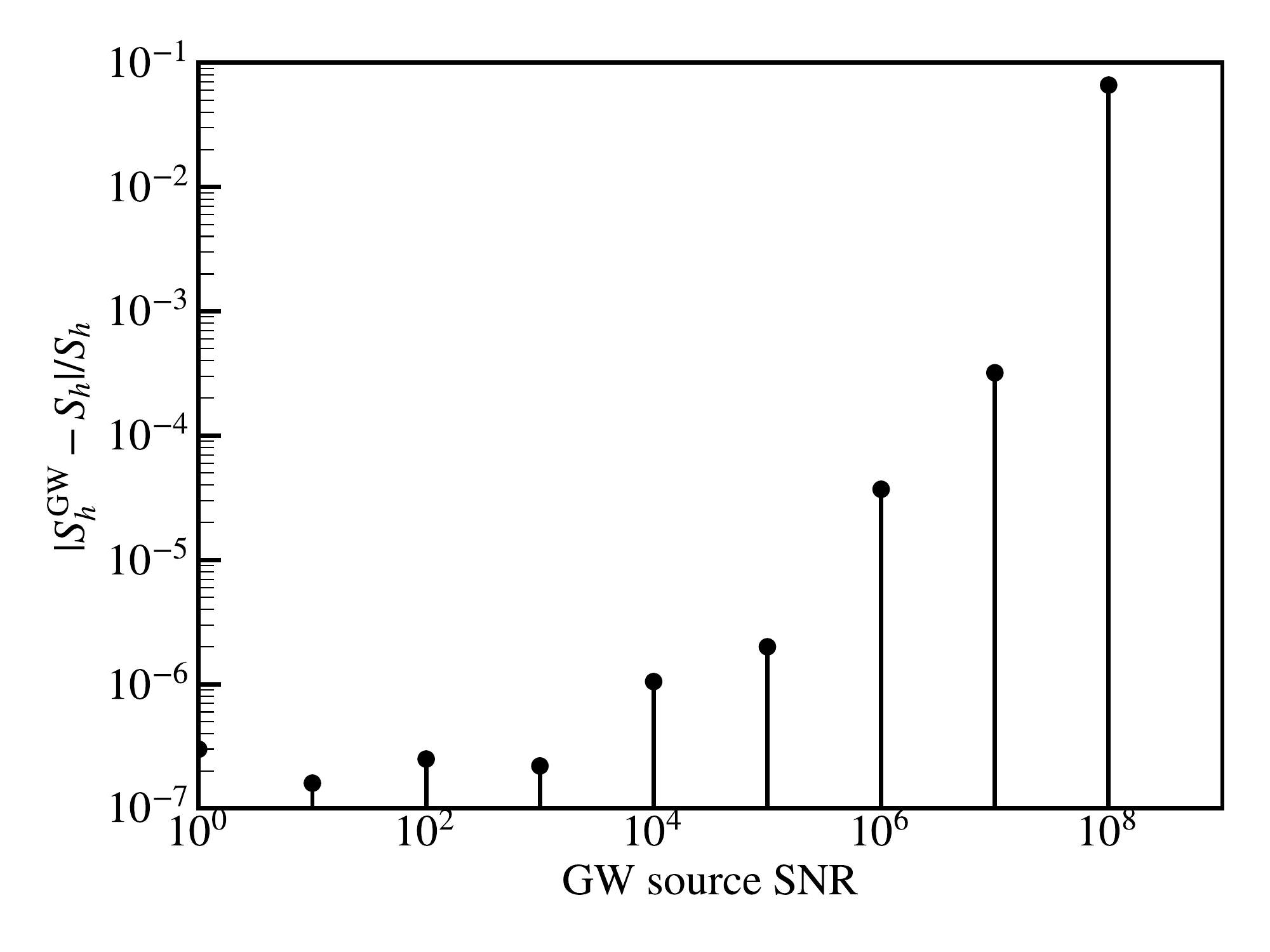}%
\caption{\label{fig:gw_impact}Impact of a 1 mHz monochromatic GW source onto the aPCI sensitivity: relative difference of the sensitivity obtained with the presence of the source relative to the source-free sensitivity.}
\end{figure}

For each measurement set containing the GW signal, we build the data matrix, compute its PCA decomposition, and derive the corresponding averaged sensitivity that we denote by $S^{\mathrm{GW}}_{h}$. We compare these sensitivities to the signal-free sensitivity (derived with data containing only noise) denoted by $S_{h}$ by computing their relative error maximized over the frequency range:
\begin{equation}
    \gamma = \underset{f}{\mathrm{max}} \left\{\frac{\left| S^{\mathrm{GW}}_{h}(f) - S_{h}(f)\right|}{S_{h}(f)}\right\}.
\end{equation}
We report the results in Fig.~\ref{fig:gw_impact} where we plot $\gamma$ as a function of the source SNR. The relative error in sensitivity remains below 1~\% for SNR values below or equal to $10^7$. Beyond $\mathrm{SNR} = 10^8$, the error becomes significant, reflecting the fact that the PCA decomposition starts to be affected by the presence of the injected signal. This demonstrates the robustness of the method in the presence of deterministic signals, as the chosen SNRs are extremely large relative to typical GW sources, especially given the short observation time we consider. As a comparison, the total GW SNR accumulated over the mission duration should be of order $\sim 10^{4}$, which is way below the threshold where it starts to affect the PCA decomposition.

\section{\label{sec:discussion}Discussion}

We outlined a method designed to process space-borne interferometer telemetry and minimize laser frequency noise for data analysis purposes. This method, aPCI, is completely blind: it does not rely on any model describing the measurements. The only implicit assumption is that there exists some unknown linear relationship over short times between the various measurements. We begin by building a data matrix by stacking integer-shifted versions of all single-link measurements; then we compute its PCA decomposition. Projecting the data onto the components with lowest variances yields data streams that are almost free from laser noise, and are directly usable for gravitational-wave searches. We demonstrate its equivalence with Bayesian inference based on TDI. This approach is a robust, complementary alternative to existing TDI techniques which depend on a specific model of the data (the laser noise equations) and on physical parameters (the time delays). This model-independent technique could be an important tool to ensure the robustness of the data analysis in missions such as LISA.

We demonstrated the effectiveness of the method to remove laser noise using simulated LISA single-link measurements, assuming fixed armlengths. We showed that the accumulated GW sensitivity of the 6 lowest-variance PCA projections is always effectively equal to the sensitivity of first-generation Michelson TDI combinations, and that the major part of the SNR concentrates in the 2 most sensitive channels.

This work lays the foundation for future investigations to build a fully data-driven approach to space-based interferometer data-analysis. Many aspects remain to be studied in more detail. First, the ability of the method to process measurements obtained with time-varying armlengths and other realistic details should be tested. Second, even though we showed that the method is robust against loud monochromatic gravitational-wave signals, we should extend this assessment to the presence of various sources across the full sensitive bandwidth, as expected in LISA. Third, one should investigate how aPCI can be further improved towards a better compression of the GW information, using orthogonality with respect to the sensitivity. \hl{Fourth, similar to TDI, aPCI would have to face data artifacts like instrumental transients and data gaps expected in LISA-like detector measurements. We should study the impact of these perturbations on aPCI performance and test its combination with mitigation techniques such as imputation}~\cite{baghi_gravitational-wave_2019}. Finally, an important step to take is \hl{to fully develop} and demonstrate Bayesian inference analysis using this approach.

\appendix

\section{\label{sec:secondary_noises}Pathlength noises}
We provide here the analytical forumlas for the secondary noise PSDs used in this study.
The acceleration noise PSD is given by
\begin{align}
    S_{a}(f) = a_{\mathrm{TM}}^2 \left[1 + \left(\frac{f_{1}}{f}\right)^2 \right] \left[1 + \left(\frac{f}{f_2}\right)^4\right]
\end{align}
where $a_{\mathrm{TM}} = 3 \times 10^{-15}\,  \mathrm{m s^{-2}}$, $f_{1} = 4 \times 10^{-4}$ Hz and $f_2 = 8 \times 10^{-3}$ Hz.
The OMS noise PSD is
\begin{align}
    S_{\mathrm{oms}}(f)= a_{\mathrm{oms}}^2 \left[1 + \left(\frac{f_3}{f}\right)^4\right],
\end{align}
where $a_{\mathrm{oms}} = 15 \times 10^{-12} \, \mathrm{m Hz^{-1/2}}$ and $f_3 = 2 \times 10^{-3}$~mHz.
\begin{acknowledgments}
We would like to thank Tyson Littenberg and Jessica Page for their interesting feedback. We also thank Martin Staab for his thorough reading and remarks. This work is partly supported by an appointment to the NASA Postdoctoral Program at the Goddard Space Flight Center, administered by Universities Space Research Association under contract with NASA.
\end{acknowledgments}

\bibliography{library}

\end{document}